\documentclass[preprint]{aastex}

\lefthead{Cha, {\it et al.}}
\righthead{LISM in Puppis-Vela}

\begin{document}

\newcommand{\kms}{\,km\,s$^{-1}$}     
\newcommand{\vlsr}{V$_{\sc LSR}$}
\newcommand{\vhel}{V$_{\small helio}$}

\title{The Local Interstellar Medium in Puppis-Vela\altaffilmark{1}}

\author{Alexandra N. Cha\altaffilmark{2}, M. S. Sahu\altaffilmark{3,4}, 
H. Warren Moos\altaffilmark{2}, \& A. Blaauw\altaffilmark{5}}

\altaffiltext{1}{Based on observations collected at the European Southern
Observatory at La Silla, Chile}
\altaffiltext{2}{Department of Physics \& Astronomy, The Johns Hopkins University,
Baltimore, MD  21218}
\altaffiltext{3}{NASA/Goddard Space Flight Center, Code 681, Greenbelt, MD  20771}
\altaffiltext{4}{National Optical Astronomy Observatories, 950 North Cherry Avenue, Tucson, AZ~87519-4933}
\altaffiltext{5}{Kapteyn Instituut, Postbus 800, 9700 AV Groningen, 
The Netherlands}

\begin{abstract}
The first study of the local interstellar medium (LISM) toward
Puppis-Vela ($l$ = 245$\degr$ to 275$\degr$, $b$ = $-$15$\degr$ to 
+5$\degr$, d $<$ 200 pc) is presented in this paper.  A study of
the locations, sizes, and physical characteristics of local interstellar
gas, i.e. ``astronephography,'' is included, and relies upon the
improved distance measurements provided by Hipparcos parallax 
measurements.  All
spectra of more distant sight lines contain absorption features due
to intervening local gas, and more distant structures can only be
studied accurately if components due to the LISM have been isolated.
Towards this end, high resolution (R $\approx$ 95,000), 
high signal-to-noise (S/N $\sim$ 110 to 250)
\ion{Na}{1} $\lambda$$\lambda$5889.951, 5895.924 spectra of 11 nearby
stars in the direction of Puppis-Vela have been obtained
with the Coud\'e Echelle Spectrograph 
on the 1.4 meter Coud\'e Auxiliary Telescope at the European Southern 
Observatory.  
Toward Puppis-Vela, absorption due to the Local Interstellar Cloud (LIC)
was not observed, but components at three
distinct velocities were found, and the extent of the local gas 
producing the features was estimated.  The three components have the following
locations and velocities: Component A---[$l$ $\approx$ 276\degr\ to 298\degr,
$b$ $\approx$ $-$5\degr\ to +4\degr], V$_{helio}$ = +6 to +9 \kms, and
d $\sim$ 104 pc; Component B---[$l$ $\approx$ 264\degr\ to 276\degr,
$b$ $\approx$ $-$7\degr\ to +3\degr], V$_{helio}$ = +12 to +15 \kms, and
d $\sim$ 115 pc; Component C---[$l$ $\approx$ 252\degr\ to 271\degr,
$b$ $\approx$ $-$8\degr\ to $-$6\degr], V$_{helio}$ = +21 to +23 \kms, and
d $\sim$ 131 pc. The 
conclusions regarding the ultraviolet
spectrum of $\gamma$$^2$ Vel ($l$ = 263$\degr$, $b$ = $-$8$\degr$, 
d = 258$\pm$35 pc) presented by Fitzpatrick \& Spitzer (1994)
were re-examined in light of this new LISM data, and the ambiguity
in their conclusions about several absorption components is
resolved.
The stars in Puppis-Vela flank 
the region of the apparent extension of the Local Bubble (or Cavity) 
known as the $\beta$
CMa tunnel, and measurements of the \ion{Na}{1} column density towards the
sample
stars have been used to modify existing estimates of the extent of the tunnel.
A compilation of all existing \ion{Na}{1} observations of $<$ 200 pc
sight lines around the tunnel reveal that 
low column densities have been exclusively detected 
within $l$ $\approx$ 210\degr\ to 250\degr, and $b$ $\approx$ 
$-$21\degr\ to $-$9\degr.  Near the Galactic plane, at latitudes
$-$10\degr\ $<$ $b$ $<$ 0\degr\ and d $\lesssim$ 150 pc, the tunnel 
is confined to l $<$ 270\degr, a lower longitude than was previously reported.

\end{abstract}


\section{Introduction}
We view the Universe through local interstellar gas and this 
gaseous material in the immediate vicinity of the Sun affects all
absorption line studies of more distant Galactic lines of sight.  

The Local Interstellar Medium (LISM) in the direction of Puppis-Vela
[$l$ = 250\degr\ to 275\degr, $b$ = $-$15\degr\ to +5\degr]
was selected for a \ion{Na}{1} absorption line study 
to determine the kinematics of large-scale structures in the 
interstellar medium (ISM) toward Puppis-Vela, and was initiated by 
Sahu \& Blaauw (1994).
Within $\sim$ 2 kpc in this direction, three large structures have been 
identified 
(1) the Vela Molecular Ridge at $\sim$ 1 kpc (Murphy \& May 1991),
(2) the IRAS Vela Shell at $\sim$ 450~pc (Sahu, 1992)
and (3) the 36$^\circ$-diameter,
H$\alpha$ emitting Gum Nebula.  \ion{Na}{1} absorption caused by 
local (d $<$ 200 pc) interstellar gas is present in 
spectra toward the distant 
background stars embedded in and beyond these extended objects. The absorption
components arising within 200~pc must
first be identified and understood before attempting to study the kinematics
of more distant ISM structures.  This work is the first attempt to
three dimensionally (l,b,d) place components of interstellar gas in a
localized region on the sky using optical absorption spectroscopy, and is 
preliminary to mapping clouds and extended structures at larger distances.
The only interstellar clouds that have been mapped three dimensionally
are the Local Interstellar Cloud (LIC) and G-cloud
(Linsky {\it et al.} 2000) using principally
H column densities from Hubble Space Telescope (HST) Space Telescope 
Imaging Spectrograph (STIS) and Goddard High Resolution Spectrograph (GHRS)
spectra and the assumption that the clouds have constant density.  
For the Puppis-Vela sightlines discussed here, the technique of Linsky {\it
et al.} cannot be applied since the Sun is not embedded in the gas
under study and there is no evidence to allow the assumption
that the lines of sight have 
constant density.  Distances to target stars derived from Hipparcos 
parallax measurements allow limits to be placed on the extent of
interstellar gas components, the first step in mapping the interstellar
clouds or understanding the ``astronephography'' (Linsky {\it et al.} 2000)
of the region.  With multiple sight lines at
various distances, recurring velocity features may be identified and used as
a basis for defining the locations, sizes, and characteristics of interstellar
gas.

Little is known about the LISM in Puppis-Vela
within $\sim$ 200~pc. In total, absorption line spectra of only five 
Puppis-Vela lines of sight
have been published (Crawford 1991; Dunkin \& Crawford 1999;
Ferlet {\it et al.} 1985 and Welsh {\it et al.} 1994).  There has be
no localized, systematic study of the local gas toward Puppis-Vela
until now.  One implication of this lack of information
is that LISM data must be obtained from all sky surveys (Welsh {it et
al.} 1998, G\'enova {it et al.} 1997) whose velocity and column density
generalizations may not be indicative of the nature of LISM gas in a 
specific direction.  The identification of local velocity components
toward Puppis-Vela performed here will allow local absorption features to be 
distinguished from more distant absorption features in any subsequent study
of the ISM in Puppis-Vela.  Questions posed by Puppis-Vela ISM studies that are
addressed here include:  What is the contribution of the
LIC in this direction? Does the Local Bubble (or Cavity)
extend beyond $\sim$ 70~pc in this direction? 
Are there any neutral velocity components in the LISM in Puppis-Vela?

We investigate the kinematics and structure of the LISM gas 
in Puppis-Vela using \ion{Na}{1} absorption line data of 11 stars
with Hipparcos-based distances $<$ 200~pc, in conjunction with data for 
five previously studied lines of sight. The \ion{Na}{1} spectra presented
here form a subset of a larger sample containing absorption spectra
toward $\sim$ 75 more distant stars in Puppis-Vela, which corroborate
the conclusions made here, and will be presented
in a subsequent paper.

\section {Components of the LISM}  

It is helpful to review the structures contained in the LISM
to more fully understand the Puppis-Vela lines of sight.  The
three known components of the LISM include the Local Interstellar Cloud
(LIC), the G cloud, and the Local Bubble.
The Sun is moving 
through a warm (T $\sim$ 7000~K), low density
(n(\ion{H}{1}) $\sim$ 0.1cm$^{-3}$), partially ionized interstellar cloud 
termed the LIC (Lallement \& Ferlet 1997, 
Lallement \& Bertin 1992). 
Figure~1 shows a schematic diagram of the components of the LISM as viewed
from above the plane of the Galaxy.  The LIC
is observed in projection toward most but not all nearby stars, and extends 
approximately 3 -- 8~pc. The LIC is the only interstellar cloud where 
{\it in situ} measurements of interstellar dust grains (Baguhl {\it et al.}
 1996) and interstellar gas (Witte {\it et al.} 1992)
have been performed through
{\it Ulysses} \& {\it Galileo} spacecraft observations.  Models of 
the LIC (Redfield \& Linsky 1999) place the Sun just inside the
LIC, in the direction of the Galactic Center and toward the 
North Galactic Pole. The Solar Wind is carving out a cavity in the LIC
which is slightly elongated in the direction of relative motion of the Sun
in the LIC. In the direction of the Galactic Center, another cloud termed
the G cloud is seen. This cloud is colder (T $\sim$ 5400~K) (Linsky \& Wood
1996) and is approaching the Sun at about 29 \kms. From Figure 1, it is clear
that the G cloud is not expected to be seen in the Puppis-Vela direction. 
Since the Sun is within the LIC, interstellar sight lines in all directions
contain absorption from gas associated with the LIC, but these features
are not always strong enough to be detected in \ion{Na}{1} absorption.
Specifically toward Puppis-Vela, very low \ion{Na}{1} column densities of 
$\sim$ 10$^8$ to 10$^9$ cm$^{-2}$ can be estimated using the empirical
N(H)/N(\ion{Na}{1}) formula given by Ferlet {\it et al.} (1985), and low 
color excesses, E(B$-$V) $\le$ 0.06, are detected toward d $<$ 200 pc
Puppis-Vela sight lines.  Very low electron column densities (n$_e$ $\lesssim$
0.03) are also deduced from the dispersion measure of three pulsars within 
200 pc (Toscano {\it et al.} 1999).

\notetoeditor{Please place Figure 1 here.}

Surrounding the Sun, the LIC, and the G cloud is the Local Bubble (or Cavity).
The Local Bubble is an irregularly shaped region that radially extends 
approximately 70 pc from the Sun (Welsh {\it et al.} 1998).
The Local Bubble appears to protrude
toward $\beta$ CMa [$l$ = 226\degr, $b$ = $-$14\degr], forming an 
extention called the $\beta$ CMa tunnel and borders the sight lines to
Puppis-Vela.  The extent of the 
tunnel is not well determined, and is 
discussed in detail in \S6.3.
It is not yet known whether the Local Bubble is a bounded region
created from a cataclysmic event such as a supernova explosion, or if
the Local Bubble is an intercloud region, isolated by the boundaries of
neighboring structures (Snowden {\it et al.} 1990).
The majority of the volume of the Local Bubble is filled with 
hot X-ray emitting plasma with a characteristic temperature of 
T $\sim$ 10$^6$K, while most of the mass in the Local Bubble is cool
and diffuse (T $\sim$ 100~K).  Typical densities in the Local Bubble
range between $\sim$ 0.002 -- 500 cm$^{-3}$ (for a recent review on this
subject see Breitschwerdt (1998)). 

\section{Using \ion{Na}{1} to study the LISM} 

There are advantages and disadvantages
associated with using \ion{Na}{1} to study the LISM. The main advantage
of \ion{Na}{1} is that it is possible to perform high resolution
(R $\sim$ 100,000) surveys with ground based telescopes.  \ion{Na}{1} is
a tracer of cold, high column density gas since it has a low ionization
potential of 5.14 eV.  On the other hand, the disadvantage of observing 
\ion{Na}{1} is that it is a trace ion in the ISM.  \ion{Na}{1}
is well suited for mapping the boundaries of hot plasma in the Local Bubble
since low column densities of \ion{Na}{1} are expected within the Local Bubble
and higher column densities are detected without.  Additionally, cold
clouds are embedded in the hot Local Bubble, and these clouds may be
located and mapped with \ion{Na}{1}.  Estimates on the extent, distance,
velocity and \ion{Na}{1} column density
of previously undocumented components containing neutral gas are presented
in \S6.  

The most abundant species in the LISM is \ion{H}{1}; however, \ion{H}{1}
is more difficult to observe since space based instruments must be 
employed to measure its column density.  
There are two empirical methods by which column densities 
of \ion{H}{1} may be estimated: the ratio of N(\ion{H}{1}):E(B$-$V)
(Bohlin {\it et al.} 1978, Diplas \& Savage 1994), and the ratio
of N(H):N(\ion{Na}{1}) (Hobbs 1974a, 1974b, 1976, Stokes 1978, and
Ferlet {\it et al.} 1985).  Neither of these two methods of empirically
calculating N(\ion{H}{1}) are suitable for the study of the LISM 
toward Puppis-Vela.

The first method of calculating atomic hydrogen column densities given 
the color
excesses of stars does not produce accurate results for sight lines 
with very low reddening since the uncertainties associated with
the MK spectral type are large compared to the observed values of (B$-$V).
The second method of determining the distribution of hydrogen estimates
N(\ion{H}{1} + H$_2$) given column densities
of \ion{Na}{1} is also inaccurate for regions of low column density.  
The studies mentioned above contain few
data points for N(\ion{Na}{1}) $\lesssim$ 10$^{11}$ cm$^{-2}$. 
Nearly 80\% of the local Puppis-Vela LISM sight lines contain absorption 
components with column densities below 10$^{11}$ cm$^{-2}$, the
regime where the N(H) to N(\ion{Na}{1}) empirical relationship
has not been shown to apply.  In addition to the lack of data points 
in the low column
density limit, Welty {\it et al.} (1994) have also noted that several
of the low column density data points are incorrect.  The large
uncertainty in N(H):N(\ion{Na}{1}) in the low column density limit makes it 
useful for only order of magnitude estimates.  As a point of reference,
however, direct \ion{H}{1} observations along several lines of sight
in Puppis-Vela at d $\sim$ 100 pc exhibit log N(\ion{H}{1}) $\approx$
18.1 to 20.1 cm$^{-2}$ (Table 5.1, Dring 1997).

\section{Observations and Data Reduction}
Na D$_1$ and D$_2$ $\lambda\lambda$5889.951, 5895.924 spectra 
for 11 B-type stars within 200 pc in the direction of the IRAS Vela
Shell ($l$,$b$) $\approx$ (263\degr,8\degr) were obtained by
M. S. Sahu and A. Blaauw using the Coud\'e Echelle Spectrograph 
(CES) on the 1.4 meter Coud\'e Auxiliary Telescope (CAT) 
at the European Southern Observatory.  The observations were made
both on site in La Silla and remotely from Garching in January 1993
(HD\,76805) and January 1994.  
The Long Camera and the UV-coated
Ford Aerospace/Loral 2048 $\times$ 2048 CCD (\#27) were used for all
observations.  The CCD's pixel size was 15 $\mu$m $\times$ 15 $\mu$m, and
had a low dark
current of 3 e$^-$/pixel/hour, a low readout noise ($\sim$ 6e$^-$ rms), and
few apparent defects.  The net efficiency of the system is 3.8\% at
5400 \AA\ and 4.6\% at 6450 \AA\ (Pasquini {\it et al.} 1992).

Table~1 contains general stellar information for the 11 stars observed.
To supplement the sample, we searched the literature and found five
additional stars with \ion{Na}{1} column density and velocity
measurements in the direction towards Puppis-Vela.  
The following information is listed for
each star: HD number, Galactic position (longitude and latitude), MK
spectral type classification, visual magnitude, observed Johnson
photometric colors from the Tycho catalogue,
calculated B--V color excess, distance obtained using Hipparcos
trigonometric parallax data, heliocentric radial velocity ($V_{rad}$), 
projected 
rotational velocities ($v\,sin\,i$), and references for
the MK spectral type and $V_{rad}$, respectively.  

\notetoeditor{Please place Table 1 here.}

A standard data reduction procedure was followed that first included
bias subtraction and flat-fielding of the science spectra.
A thorium-argon arc lamp was the wavelength calibration source, and
the stability of the CES yielded unchanging calibration exposures throughout
each night.  The calibration spectra were used in conjunction with 
the Th-Ar line list by Willmarth (1987) to visually identify the
emission lines, fit a second degree polynomial to the resulting pixel vs.
wavelength arrays, and thereby convert our absorption line spectra from pixel
to wavelength space.  The absolute wavelength solutions had rms
variations of 0.7 \kms\ or 0.014 \AA\ for the 1993 spectrum and 
1.1 \kms\ or 0.023 \AA\ for the 1994 spectra.  The instrumental resolution
determined using the FWHM of the thorium lines
for both observing runs was 3.1 \kms, equivalent to 
$\lambda/\Delta\lambda$ $\approx$ 95,000.

In addition to the interstellar \ion{Na}{1} absorption features,
the raw spectra were contaminated by 
many telluric absorption lines which are
abundant at wavelengths surrounding the \ion{Na}{1} D doublet.
Multiple observations of 
$\alpha$ Vir (HD\,116658) were made since it is a nearby star with a 
high rotational velocity (v sin i = 159 \kms, Hoffleit \& Jaschek 1982) and 
little interstellar \ion{Na}{1} absorption.  
These spectra were used as templates
to remove telluric absorption lines contaminating the spectra.
To normalize the
template spectra, the continuua were fit with cubic splines then divided
by the fits.  The optical depths of the atmospheric absorption lines 
in $\alpha$ Vir were adjusted by a multiplicative factor for each star
so that the strength of the well separated telluric lines
at wavelengths of 5883$-$5901 \AA\ matched those in each object spectrum.
Next, the object spectra were divided by the scaled $\alpha$ Vir template
spectra to eliminate telluric absorption lines.  The observed spectra 
were continuum normalized by fitting a cubic spline and dividing
the object spectrum by the fit.  
The resulting spectra had signal-to-noise ratios of 110 to 250.

\section{Presentation of the Data} 
To fit the spectral lines and infer physical properties about the
interstellar velocity components along each line of sight, we used the profile 
fitting method and software developed by Welty, Hobbs, \& Kulkarni (1994).
This technique assumes that the components have Maxwellian velocity
distributions.  Each line was fit with the fewest number of 
components necessary, and additional components either increased the 
$\chi$$^2$ statistic or required that one or more components have unphysical
b values.  
The parameters describing each component were adjusted by
an iterative, non-linear, least-square method to achieve an rms 
deviation of the absorption line fit that was comparable to the rms
deviation in the stellar continuum, and reduced the value of the
$\chi$$^2$ statistic.  

The \ion{Na}{1} spectra of the 11 stars observed are displayed in
Figure~2.  For each sight line, normalized intensity is plotted versus 
heliocentric velocity for both the D$_2$ and D$_1$ lines.  Filled
circles indicate the data points, 
the best-fit models are plotted with solid lines, and
dashed lines trace the individual Gaussian components.  Beneath the
D$_1$ spectra are the residuals to the fits for the D$_2$ lines.  
For the spectra of 
HD\,72232 and HD\,79416, the y-axes in the residual plots range 
from $-$0.1 to 0.1 in units of normalized intensity, whereas for the
rest of the spectra, the y-axes of the residual plots range from
$-$0.015 to 0.015.  Upwards arrows beneath the D$_2$ lines are located
at the projected LIC velocity.  No LIC components are 
revealed in the spectra, as expected (see \S2).

\notetoeditor{Please place Figure 2 here.}

Table~2 contains numerical data associated with the model fits and
includes: (1) the HD number,
(2) central heliocentric velocity, (3) equivalent width, (4)
Doppler b parameter, (5) logarithm of the \ion{Na}{1} column density,
(6) signal-to-noise ratio, and (7) the heliocentric to LSR velocity 
conversion factor.
Uncertainties were estimated by comparing the
measured \ion{Na}{1} D$_2$ and D$_1$ profile fit parameters.  If there were
no sources of error, these values would be identical since each pair
of doublet lines arise due to absorption by the same \ion{Na}{1} gas.
The maximum difference in measured velocities between a D$_2$ versus D$_1$
component is $\pm$0.5 \kms, and is consistent
with the calculated rms velocity dispersion.  The uncertainties associated
with the column density and b value are generally $<$ 10\% and a quality
estimate is listed for each spectral line in Table~2.

\notetoeditor{Please place Table 2 here.}

\section{Discussion}

\subsection{Identification of Three Velocity Components}
Along the eleven sight lines toward early type stars located 
within 250\degr\ $<$ $l$ $<$ 299\degr\ and
$-$8\degr\ $<$ $b$ $<$ +4\degr, \ion{Na}{1} gas is primarily found 
in one of three distinct velocity ranges (to an accuracy
of $\sim$ 1 \kms) .  Five additional sight lines with 
\ion{Na}{1} kinematic information were found in the literature.  
General stellar information for these stars is listed in Table~3.  
Information about the \ion{Na}{1} components observed is
in Table~4 and includes: (1) HD number, (2) central 
heliocentric velocity, (3) Doppler b parameter, 
(4) logarithm of the \ion{Na}{1} column density, 
(5) velocity resolution at which the observation was made, and (6) a reference
for the data.

\notetoeditor{Please place Tables 3 and 4 here.}

Although the sixteen lines of sight are
spread over $\sim$ 50\degr\ in Galactic longitude, the absorption
components do not appear at random velocities.
The majority of the velocity components can be associated with one 
of the following
velocity ranges: (A) +6 \kms\ $<$ V$_{helio}$ $<$ +9 \kms, (B) +12 \kms\
$<$ V$_{helio}$ $<$ +15 \kms, and (C) +21 \kms\ $<$ V$_{helio}$ $<$ 23 \kms.
The Galactic coordinates of the sixteen lines of sight have been plotted in
Figure~3.  Filled triangles
indicate the locations of stars whose we present new \ion{Na}{1} data,
while open triangles pinpoint the locations of stars for which data are from 
the literature.  To illustrate the presence of absorption
components at the three common velocities described, lines of sight with
absorption at velocities A, B, and C have been encircled.
No distinction has been made regarding the source of the data in Figures~3b-3d.
The symbol sizes indicate
the relative \ion{Na}{1} column density detected.  Small symbols correspond
to 10.0 $<$ log[N(\ion{Na}{1})] $<$ 10.4, medium symbols
denote 10.4 $<$ log[N(\ion{Na}{1})] $<$ 11.1, and large symbols
represent column densities of 11.7 $<$ log[N(\ion{Na}{1})] $<$ 12.6.

\notetoeditor{Please place Figure 3 here.}

Analysis of the distribution of data points in Figure~3
reveals that velocity components A, B, and C are located at distinct 
regions in $l$ and $b$.  The size of the data points, which corresponds to
the \ion{Na}{1} column density, reveals that higher column densities of
gas at velocities A and B exist in this region of the LISM, whereas the
sight lines with absorption at velocity C have lower column densities.
Comparing the locations where each of the components are 
detected, the velocities
of the components in the Local Standard of Rest
decrease with increasing Galactic longitude.  General Galactic rotation
also follows this trend, but the Galactic rotation velocities 
calculated at d = 150 pc,
assuming a galactocentric distance of 8.5 kpc, and
a local circular speed of 220 \kms\ in the solar neighborhood (Mihalas
\& Binney 1981), do not
match up with the velocities of \ion{Na}{1} absorption features
observed in the LISM.  The sightlines to stars at d $>$ 200 pc
confirm the locations of local gas absorption for components B and C,
whereas the region 280$\degr$ $<$ $l$ $<$ 300$\degr$, where component A
is detected, is outside of the 
sample area.  Also from the spectra of the more distant sight lines, 
absorption features at the velocity of Component B were
additionally observed along sightlines with $b$ $\gtrsim$ $-$10$\degr$ and 
$l$ $\gtrsim$ 260$\degr$, and gas with velocities coincident with 
Component C was observed up to the Galactic plane ($b$ $\lesssim$ 0$\degr$).
 
In Figure 4 the Puppis-Vela region is viewed from above the Galactic plane
and contains the same data points as Figure~3.  Here,
distance increases radially outward and galactic longitude
increases counterclockwise.  The symbol sizes are defined identically
as in Figure~3, and the same Galactic latitude range is applied: $-$10\degr $<$
$b$ $<$ +5\degr.  The symbols in Figure~4 indicate the 
velocity of the absorption component, and have been placed at the appropriate 
distance and galactic
longitude of the target star.  From this perspective, the locations of 
components with similar velocities are also seen to be grouped together, 
and not randomly distributed.  Arcs have been overplotted to highlight
the maximum distance and minimum extent in Galactic longitude of the
front edge of a gas component at a particular velocity.
The arcs have been labled with ``A,'' ``B,'' or ``C''
to match the velocity ranges defined above and on the plot.  
Note that the arc labeled ``C'' is dashed in the center to
indicate where its presence is uncertain according to the spectra presented
here, but when spectra from more distant stars are included, gas with
velocity C is seen throughout the entire Galactic longitude range.
 
The locations and characteristics of the three components are
summarized in Table~5.  For each component, Table~5 lists
the minimum extent in Galactic longitude and latitude where the gas
is observed, the maximum distance to the gas, the heliocentric velocity
range of the gas, and the spread in \ion{Na}{1} column densities for 
spectral lines within the given velocity range.  Three stars (HD\,61831,
HD\,65575, and HD\,74195) have been
omitted from Figure 4 for reasons discussed in \S6.3.
  
\notetoeditor{Please place Figure 4 here.}

Sight lines either pass completely through the component, or partially 
penetrate the gas.  Using the distribution of the measured column densities
towards the target stars of known distances, it is possible to deduce 
limits on the
distances to each gas component.  Refining these distance estimates would
necessarily require more observations ($\sim$ 18 additional sight lines
exist) to obtain a finer distribution of
observations so that the boundaries of the gas and the density distribution
of the gas within the components could be distinguished.  Specific 
characteristics of the gas at velocities A, B, and C are suggested below.

The three components of gas observed in the LISM have unique characteristics.  
Sight lines with velocity A gas
exhibit rather high column densities (large symbols)
along three out of four sight lines.
These three high column density sight lines (\#1, 3, 8) are located at 
Galactic latitudes of
$-$4.9\degr\ $<$ $b$ $<$ $-$3.0\degr, while the fourth target whose spectrum
exhibits gas with velocity A is at $b$ = +3.8\degr.
Since the latitude of the latter target, HD\,106490, is on average 8\degr\ away
from the others, a smaller column density of gas
at \vhel\ = +6 to +9 \kms\ exists
above the galactic plane towards HD\,106490 (\#2).  The absorbing material is
confined to a distance $<$ 100 pc since the column density of
\ion{Na}{1} does not increase when more distant sight lines are observed.

Sight lines with components at velocity B tend to have moderate
to high column densities at larger distances.  Toward the stars
HD\,79416 (\#9) and HD\,72232 (\#10) we find log N(\ion{Na}{1})~$\sim$~11.8,
whereas the remainder of the sight lines with absorption at velocity 
B ranged from log N(\ion{Na}{1}) = 10.28 to 11.08.
Towards one of the seven stars with spectral features at velocity B, HD\,74560
has a low column density (small symbol), yet it is physically 
located between several stars that have substantially higher column densities.
This variation in the magnitude of the column densities suggests
that the velocity B gas is patchy, and inhomogeneous.  
The high column density along the line of sight towards HD\,74195 (\#8),
observed by Welsh {\it et al.} (1994), reveals gas at velocity B, however
the Doppler parameter given, b = 0.3$\pm$0.1 \kms\, is small compared to 
the 3.6 \kms\ velocity resolution, so this data point has been
omitted from our analysis.

There are two concentrations of sight lines with velocity C gas: one
at $l$ $\approx$ 270\degr\ and one at $l$ $\approx$ 252\degr.  Low column 
densities (small symbols) are observed towards the stars grouped at 
$l$ $\approx$ 270\degr\ while moderate column densities (medium symbols)
are seen toward $l$ $\approx$ 252\degr.  Stars HD\,74146 (\#5), HD\,74071
(\#6), and HD\,74560 (\#7) may be located near the edge of the component or
the component may be very diffuse
considering the low values of N(\ion{Na}{1}) detected.

Because of the absence of observed sight lines with d $<$ 200 pc, 
the arc tracing the estimated boundary of
velocity C gas is dashed in the middle.  Three nearby stars are located behind
the dashed portion of the arc but do not exhibit absorption by gas at
velocity C in their \ion{Na}{1} spectra.  As noted above, it is likely
that the gas is patchy and inhomogeneous, which would then account for
the observations.  The presence of component C in the spectra of the
longer sight lines in the sample is also intermittant, also indicating that
the component C gas is patchy.  One of these stars, HD\,79416 (\#10), 
is located above the
Galactic plane at $b$ = +3.3\degr, while all of the stars revealing velocity
C gas have $b$ $\le$ $-$5.9\degr.  The spectra from the more
distant stars in our sample include component C gas 
for $b$ $\lesssim$ 0$\degr$ and for the same extent in $l$.

With the components of neutral \ion{Na}{1} gas arising in the first
200 pc of the ISM toward Puppis-Vela identified, any subsequent 
spectroscopic study of the neutral ISM in this direction can distinguish
between local gas and more distant gas.  The presence of absorption 
components at velocities A, B, or C in the spectra of future studies will
not be confused with absorption due to the myriad other structures that
exist toward Puppis-Vela.  The three dimensional information about the
components--the extent of the gas as projected on the sky, and the limits
on the distance to the gas--serve as a basis of estimating the typical
velocities and column densities present in the LISM.

In general, the LISM toward Puppis-Vela has low positive velocities ranging
from +6 \kms\ to +23 \kms.  Absorbing gas at these velocities is
located at distances closer than $\sim$ 104 pc, and the \ion{Na}{1}
column densities detected range from log N = 10.2 cm$^{-2}$ to 11.9 cm$^{-2}$
indicating that the cold gas is clumpy or patchy, and not uniformly
distributed.

\subsection{Using LISM Component Data--An Example}

To put our results in perspective, we searched the literature for
UV interstellar absorption line studies in the direction of Puppis-Vela
in order to demonstrate that knowing the properties of the LISM
gas in a particular region of the sky is extremely valuable for 
the interpretation of longer lines of sight.
The only recent study has been toward $\gamma$$^2$ Velorum by
Fitzpatrick \& Spitzer (1994; hereafter FS94).
Their {\it Hubble Space Telescope (HST)}/Goddard High-Resolution Spectrograph
(GHRS) observations have a velocity resolution similar to our 
\ion{Na}{1} data (3.1 \kms\ versus 3.5 \kms\ for the GHRS data)
facilitating a comparison between 
the two datasets.  $\gamma$$^2$ Velorum is located at 
[($l$ = 263\degr, $b$ = -8\degr); $d$ $\approx$ 260 pc] so it is
likely that some of the components observed in the GHRS spectra
arise in the LISM. FS94 detect seven components as described in
Table~6.  Comparing the
velocities of the FS94 features with the three velocity components we detect
in the LISM, their component 6 corresponds to 
our component C and their components 4 and 5 correspond to our component B. 

The neutral species detected with GHRS include \ion{S}{1} and \ion{C}{1}.
FS94 fit the \ion{S}{1} spectrum with component 4, and the \ion{C}{1}
feature with components 2 and 4.  In addition to these neutral species,
we also include a \ion{Na}{1} spectrum of $\gamma$$^2$ Vel that
was obtained at the same time and with the same instruments as the LISM
data toward Puppis-Vela described
in \S4.  The \ion{Na}{1} D$_2$ and D$_1$ spectra of $\gamma$$^2$ Vel are
shown in Figure~5,
and the parameters of the fit are listed in Table~7.
\ion{Na}{1} absorption is seen at velocities corresponding to FS94
components 2, 4, and 6.  Comparing the \ion{Na}{1} data for the
$\gamma$$^2$ Vel line of sight with the distribution of gas in the
foreground, we may conclude that component 2 arises in gas located
beyond 200 pc, yet less than 260 pc---the distance to $\gamma$$^2$ Vel.
Both components 4 and 6 were detected in the LISM at distances $<$ 200 pc.

The presence of component 4 is complicated, however, 
by the fact that it appears in the spectra of all of the species detected
in FS94, both neutral and highly ionized, and in our \ion{Na}{1} spectrum.
In addition, FS94 note that a {\it Copernicus} spectrum toward
$\gamma$$^2$ Vel reveals H$_2$ absorption which may be assigned to
either component 4 or a blend of components 2 and 5.  Although FS94
assign \ion{H}{2} region origins to component 4 absorption because of
the presence of species such as \ion{C}{4} and \ion{Si}{4}, it is 
difficult to reconcile neutrals and molecules 
coexisting with such highly ionized species.  

FS94 propose that the \ion{H}{2} region containing the component 4 gas is the
$\sim$ 40 pc surrounding a
$\sim$ 60 pc-radius stellar wind blown bubble centered on $\gamma$$^2$ Vel.
We suggest
that there are two distinct regions of gas with velocities matching
component 4.  First, the ionized region extending ~100 pc from the star 
described in FS94 contains ionized gas at V$_{helio}$ $\approx$ 13 \kms.
Second, the neutral gas absorption along the line of sight 
to $\gamma$$^2$ Vel at V$_{helio}$ $\approx$ 13 \kms\ occurs at d $\approx$
115 pc, corresponding with our component B.  If the H$_2$ spectrum is fit by
component 4, then it is consistent to place the H$_2$ gas with the \ion{Na}{1}
in component B, thereby providing a unique example of H$_2$ gas near the
edge, or possibly within, the Local Bubble.

Not only is the mapping of the nearby gas essential in the
characterization of the LISM itself, this data clarified the complex
absorption signature along the $\gamma$$^2$ Vel line of sight, and
should prove useful for other Puppis-Vela sight lines.
Caution must be observed, however, since it has been
shown here and in other papers (Watson \& Meyer (1996); Frail {\it et al.} 
(1994); Lauroesch {\it et al.} (1998, 1999)) that interstellar \ion{Na}{1} is 
patchy, as evidenced by the subparsec-scale structures that appear to be
ubiquitous in the diffuse ISM, but this focused study will detect most
components specific to the Puppis-Vela LISM which previous all-sky
local gas surveys ignored.

\subsection{LISM Absorption Components at Peculiar Velocities}

Of the 27 absorption features detected (excluding the ultra-high
resolution data of
Dunkin \& Crawford (1999)) along 16 lines of sight, 9 features have
velocities outside of velocities A, B, and C.   The peculiar velocities of
all but three components can be understood with simple explanations.
A total of three absorption components toward HD\,61831 and HD\,72232 
lie just outside of the velocity ranges A, B, and C.
It is likely that these
absorption features are blends of multiple narrow lines that are
unresolved at R $\sim$ 100,000, as evidenced by the many components
distinguished in the ultra-high resolution
data towards HD\,81188 cited in Table~4.  Three additional components 
toward HD\,65575 and HD\,106490 have peculiar velocities, but these stars
are located at the periphery of the sample, and it is natural that at some
point the characteristics of the gas change and components A, B, and C
cease being detected.

In total, only 3 of 27 components cannot be simply accounted for,
and they include the 
V$_{helio}$ = +132.3 \kms\ and $-$10.6 \kms\ lines toward HD\,62226 and
the V$_{helio}$ = $-$139 \kms\ line toward HD\,74146.
The spectra of HD\,62226 and HD\,74146 are plotted in Figure~6a and
5b, respectively.  For each line of sight, the \ion{Na}{1} D$_2$ absorption
is plotted at the top, the D$_1$ absorption is plotted in the middle of
the panel.  The residual to the best fit to the D$_2$ spectrum is on the 
bottom, plotted with the y-axis ranging from $-$0.015 to +0.015. 
Unlike the spectra of
the other LISM sight lines in this region, that of HD\,62226 contains three
components at three distinct velocities.  The absorption 
at V$_{helio}$ = +22.2 \kms\ is consistent with the gas observed towards
adjacent sight lines in the region, but the absorption
features at V$_{helio}$ = $-$10.6 \kms\ and +132.2 \kms\ are not observed 
towards any other stars in the area.

\notetoeditor{Please place Figure 6 here.}

Although lines of sight adjacent to HD\,62226 were also observed, namely
HD\,61878, HD\,64503, and HD\,61831, the spectra toward these stars
do not show absorption at either V$_{helio}$ = $-$10.6 \kms\ or +132.2 \kms.
Because the sight line towards HD\,62226 is fortuitously flanked by
these three nearby lines of sight that have also been observed, 
it may be concluded that the gas
producing the V$_{helio}$ = $-$10.6 \kms\ and +132.2 \kms\ components
are confined to a small region projected on the sky at d $<$ 190 pc.
The V$_{helio}$ = $-$10.6 \kms\ component is very weak, but is clearly
visible in both the D$_2$ and D$_1$ spectra.  The D$_2$/D$_1$ ratio of 
equivalent widths for the +132.2 \kms\ lines equals 2.2, which is close to
the expected ratio of 2 for unsaturated \ion{Na}{1} D$_2$ and D$_1$ lines.  

The other sight line with unexplained components is that of HD\,74146 which 
also contains an absorption component at
a high velocity, but here the feature is a broad (b = 16.7 \kms) line at 
negative velocity, V$_{helio}$ = $-$139 \kms, and D$_2$/D$_1$ = 0.8.  The
breadth of the line and the low D$_2$/D$_1$ ratio are atypical of 
unsaturated interstellar lines, and it is unlikely that such a weak line
could be saturated.  Additionally,
several sight lines adjacent to
HD\,74146 were also observed (HD\,76805, HD\,74071, and HD\,74560)
though their spectra did not reveal absorption at high negative velocities.

A literature search for information regarding whether or not circumstellar
material has been observed around HD\,62226 or HD\,74146 yielded nothing.
Additionally, we attempted to use {\it International Ultraviolet Explorer}
(IUE) spectra to determine the origin of the high velocity absorption
features.  HD\,74146 was not observed with IUE, and only one high dispersion
spectrum of HD\,62226 was obtained.  The single spectrum toward HD\,62226
had a low signal-to-noise ratio (S/N $<$ 10), so the presence or absence
of high velocity features associated with characteristic 
circumstellar or interstellar absorption could not be ascertained.
With the data available, we were {\it not} able to determine whether
the high velocity features in the spectra of HD\,62226 and HD\,74146
were of circumstellar or interstellar origin.

\subsection{The $\beta$ CMa Tunnel}
In \S2 it was mentioned that the extension of the Local Bubble (or Cavity)
known as the $\beta$ CMa tunnel is adjacent to Puppis-Vela when projected
on the sky.  This tunnel of rarefied hot gas has been estimated to extend 
$\sim$250-300 pc (Sfeir {\it et al.} 1999;
Welsh 1991; Welsh, Crifo, \& Lallement 1998), and 
appears to be almost free of neutral gas.
For instance, \ion{H}{1}
column densities of N(\ion{H}{1}) $\sim$ 10$^{18}$ cm $^{-2}$ measured
toward the 153$\pm$15 pc line of sight to $\beta$ CMa are thought to arise
primarily in gas along the first $\sim$5 pc---within the LIC (Gry {\it et al.}
1985).  In this section, previous estimates of the
tunnel's extent will be reviewed and a new measurement of the size of the
tunnel, within the first 150 pc, will be presented.

After the low gas densities towards several stars
in the direction of $\beta$ CMa were noticed,
subsequent studies of the LISM probed adjacent sight lines
to ascertain the size of the tunnel.  Using
optical absorption spectroscopy to observe the \ion{Na}{1} D doublet, 
three estimates of the extent of the tunnel were made.
First, Welsh (1991) assigned minimum angular dimensions to the tunnel:
an elliptical hole located at 226\degr\ $\lesssim$ $l$ $\lesssim$ 242\degr,
$-$11\degr\ $\lesssim$ $b$ $\lesssim$ $-$20\degr, extending to a depth of
$\sim$300 pc.  Later Welsh, Crifo, \& Lallement (1998) 
revised the size of the tunnel and substantially 
widened the estimate.  They reported minimum approximate dimensions of 
the low gas density tunnel towards $\beta$ CMa to be
250 pc long by 90 pc wide.  Analysis of Figure 2 in Welsh {\it et al.}
indicates that the tunnel narrows at larger distances; specifically, 
the extent of the tunnel is 210\degr\ $\lesssim$ $l$ $\lesssim$ 277 \degr\
at d = 150 pc, while at d = 250 pc regions of higher column density
gas bound the
tunnel to within 215\degr\ $\lesssim$ $l$ $\lesssim$ 240\degr.
Sfeir {\it et al.} (1999) included the $\beta$ CMa tunnel in their
maps of the neutral gas associated with the Local Bubble, but used
only a few data points to confine the low gas density tunnel.
Their data is inconclusive regarding the depth of the tunnel and
they proposed that the tunnel is either terminated at
a distance of $\sim$ 250 pc by \ion{Na}{1} absorbing gas producing 
equivalent widths of $\sim$ 50 m\AA, or that the boundary of the tunnel
is closer but contains several perforations.

The maps illustrating the tunnel in Welsh {\it et al.} (1998) and
Sfeir {\it et al.} (1999) do not represent differences in 
the stars' Galactic latitude, however.  Both papers
show the tunnel on plots of distance versus Galactic longitude with
the stars projected onto the Galactic plane, thus obscuring information
regarding the Galactic latitude of the target stars.  Almost all of the
published \ion{Na}{1} spectral data in the direction of $\beta$ CMa
pertain to stars below the Galactic plane at $b$ $\sim$ $-$10\degr\
to $-$30\degr.  Our observations add a new dimension to the tunnel boundary
since the target stars in our sample are concentrated within 
$b$ $\sim$ $-$10\degr\ to 0\degr.

In addition to the \ion{Na}{1} column density measurements given in 
Tables~2 and 4 for stars toward Puppis-Vela, N(\ion{Na}{1}) measurements
along sight lines in and around the $\beta$ CMa tunnel were compiled.
Stellar information (as was given in Table~1) about these published 
sight lines is in Table~8, and the corresponding \ion{Na}{1} column densities
and references are in Table~9.

\notetoeditor{Please place Tables 8 and 9 here.}

The column densities toward sight lines listed in
Tables~2, 4, and 9  and projected onto the Galactic plane are illustrated
as a function of distance and Galactic longitude, in Figure~7.
The Galactic latitudes of the lines of
sight have been divided into two groups: those at $-$10\degr\ $<$ $b$ $<$ 
0\degr\ are indicated with filled symbols, while targets located at
$-$30\degr\ $<$ $b$ $<$ $-$10\degr\ are plotted using open symbols.  
Diamonds are used when precise measurements of the \ion{Na}{1}
column density are known and triangles are plotted for sight lines where 
only N(\ion{Na}{1}) upper limits have been determined.  The symbol sizes 
suggest the relative column density detected along each line of sight, with
larger symbols indicating larger column densities, as defined in the
caption to Figure~3.

The thin solid lines in Figure~7 depict 
the longitudinal angular extent of the tunnel estimated by Welsh (1991).
In view of the data shown in Figure~7, it appears that his estimate should
be expanded by $\sim$ 5\degr\ on either
edge out to a distance of $\approx$ 150 pc.
Note that all of the sight lines supporting this tunnel estimate have $b$ $<$
$-$10\degr.  So far, no observations have been made
to constrain the tunnel's extent in
latitude or length.  The expanded extent of the tunnel proposed 
by Welsh {\it et al.} (1998) (in their Figure 2) is sketched in Figure~6
with dotted lines.  With the additional lines of sight near the Galactic 
plane presented here, the Welsh {\it et al.} high longitude tunnel
boundary appears misplaced.  Either one or more of the following
reasons may account for this.  The new data reveals that 
the first $\sim$ 150 pc of the
$\beta$ CMa tunnel do not extend beyond l $\approx$ 267\degr, 
near the plane of the Galaxy ($-$10\degr\ $<$ $b$ $<$ 0\degr). 
There is no data to confirm if the tunnel does or does not extend beyond
$l$ = 267\degr\ for $-$30\degr\ $<$ $b$ $<$ $-$10\degr.  At distances
of $\sim$ 150 to 200 pc, the tunnel does not extend beyond 
$l$ $\approx$ 251\degr\
near the plane of the Galaxy, as evidenced by the filled data points at
$l$ $>$ 250\degr\ in Figure~7.  There is no observational evidence to
determine whether the $\beta$ CMa tunnel extends past $l$ $\approx$ 250\degr\
at lower Galactic latitudes, since no sight lines have been observed there.
The lower Galactic longitude boundary also appears to encompass sight lines 
with high \ion{Na}{1} column densities, and should be relocated to 
$l$ $\gtrsim$ 215\degr.  From the compilation of data on almost 40 
lines of sight, it appears that outside the Local Bubble
(d $>$ 70 pc), low neutral gas densities are to be found 
between $l$ $\approx$ 215\degr\ to 250\degr\ and $b$ $\approx$ $-$21\degr\ to
$-$9 \degr.  The arc drawn in Figure~7 at d $\sim$ 150 pc illustrates
the extent in Galactic longitude over which low column density lines of 
sight with $b$ $\lesssim$ $-$10\degr\ have been observed.  The arrows 
pointing to larger distances suggest that this bound on the $\beta$ CMa
tunnel is only a lower limit, but is dictated by the present data. 

\notetoeditor{Please place Figure 7 here.}

The most distant sight lines included in this sample 
have d $\sim$ 200 pc and $b$ $\sim$ $-$30\degr.  At this limit, the sight
lines have not
penetrated the scale height of the disk of the Galaxy (see Dame \& Thaddeus
(1994) and Dickey \& Lockman (1990)).  Studies of more
distant sight lines attempting to map the distant 
end of the tunnel could easily be confused 
by the Galactic disk's decreasing density gradient, whereby stars at large
distances and lower latitudes might exhibit lower N(\ion{Na}{1}) measurements
because of the geometry of the Galaxy, and not necessarily because of 
extensions in the Local Bubble.  From nearby interstellar \ion{Na}{1} 
measurements, there is a region of low density towards
$\beta$ CMa, but it is not clear whether the tunnel is present in the plane
of the Galaxy.  Additional observations along sight lines 0\degr\ to 
30\degr\ below the Galactic plane are needed to clarify the three dimensional
extent of the $\beta$ CMa tunnel.

\section{Conclusions}

We have presented high resolution \ion{Na}{1} spectra toward 11 
early type stars plus kinematic
data for 5 lines of sight taken from the literature to study the LISM 
in the direction of Puppis-Vela.  Additionally, we have compiled a list
of \ion{Na}{1} column density measurements made toward nearby (d $<$ 200 pc)
sight lines in the $\beta$ CMa tunnel.  Our conclusions are stated
below:

1.  Observations of \ion{Na}{1} in the LISM toward Puppis-Vela revealed
absorption at three distinct velocities with the following properties:
Component A---[$l$ $\approx$ 276\degr\ to 298\degr,
$b$ $\approx$ $-$5\degr\ to +4\degr], V$_{helio}$ = +6 to +9 \kms, and
d $\sim$ 104 pc; 
Component B---[$l$ $\approx$ 264\degr\ to 276\degr,
$b$ $\approx$ $-$7\degr\ to +3\degr], V$_{helio}$ = +12 to +15 \kms, and
d $\sim$ 115 pc; 
Component C---[$l$ $\approx$ 252\degr\ to 271\degr,
$b$ $\approx$ $-$8\degr\ to $-$6\degr], V$_{helio}$ = +21 to +23 \kms, and
d $\sim$ 131 pc.  This identification of LISM gas will enable future
studies of the Puppis-Vela ISM to separate local gas absorption from
more distant features, as was illustrated for the $\gamma$$^2$ line of
sight.  Including a distinction of the distance at which
gas exists, as well as its position on the sky, is a fundamental step in
the new field of three dimensional astronephography.

2.  The LIC is not detected in \ion{Na}{1} absorption
toward Puppis-Vela because the
column densities of LIC gas in this direction are too low, generally
$<$ 10$^{11}$ cm$^{-2}$.

3.  Low column density sight lines in the $\beta$ CMa tunnel are confined
to $l$ = 215\degr\ to 250\degr\ and $b$ = $-$21\degr\ to $-$9\degr, to
a distance of $\sim$ 150 pc.
New observations closer to the Galactic plane ($-$10\degr\ $<$ $b$ $<$
0\degr) reveal higher column densities suggesting that the tunnel 
may not extend in latitude to $b$ = 0\degr.  More observations near the
Galactic plane and at distances greater than 150 pc are needed to
ascertain the true size and location of the $\beta$ CMa tunnel.

4.  Although most of the absorption components seen in the spectra
of stars within 200 pc toward Puppis-Vela could be assigned to one of 
the three components, A, B, or C, a few components had peculiar velocities.
Toward HD\,62226 and HD\,74146 high velocity features, V$_{helio}$ =
+132.2 \kms\ and  V$_{helio}$ = $-$139 \kms\ respectively, were present.
In both cases, nearby sight lines
were also observed and contained no high velocity features.  The 
origin of the high velocity gas components, be they
circumstellar or interstellar, is unknown.

\acknowledgments

ANC and HWM acknowledge support from NASA contract NAS 5-32985 to Johns 
Hopkins University and MSS acknowledges support from a
GTO grant to the STIS IDT.  We thank F. Bruhweiler, W. Landsman, and J.
Lauroesch for helpful comments on the manuscript and E. Burgh for a
useful graphics routine.
We acknowledge use of the Simbad Database at the Centre Donn\'ees 
astronomiques de Strasbourg (http://simbad.u-strasbg.fr/Simbad).

\newpage

\begin{center}Figure Captions\end{center}

\noindent
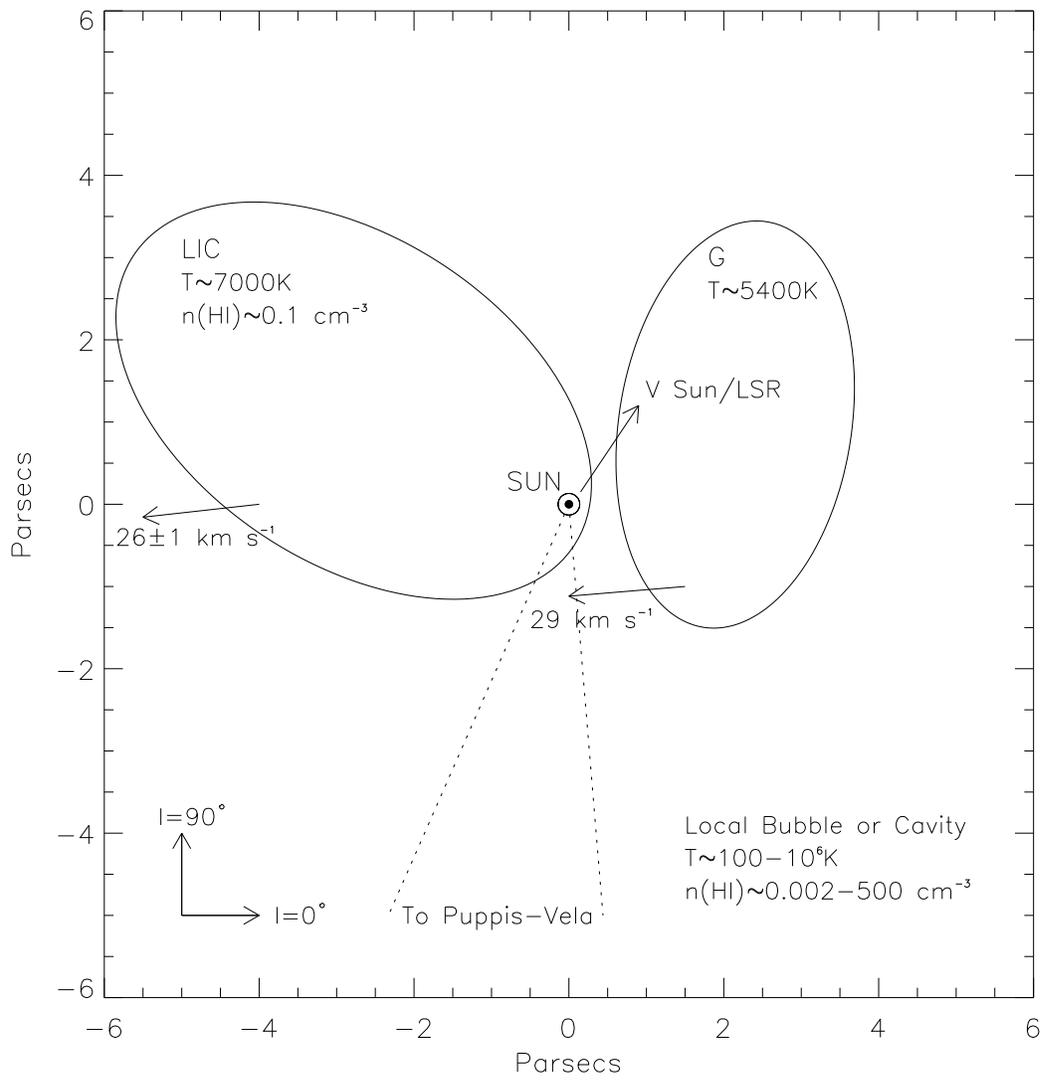
Figure~1.  A schematic diagram of the structures immediately 
surrounding the Sun.
The Sun (at center) is embedded in the LIC and near the G cloud, 
which are both within the Local Bubble.  Approximate locations, sizes,
and angles subtended by the clouds are given by Linsky {\it et al.} (1999)
and Lallement (1998).
The velocities of the LIC and G cloud shown are with
respect to the heliocentric frame of reference.  The arrow extending from the
Sun shows the direction of the motion of the Sun relative to the Local
Standard of Rest (LSR).  Dashed lines indicate
the approximate angle subtended by Puppis-Vela.

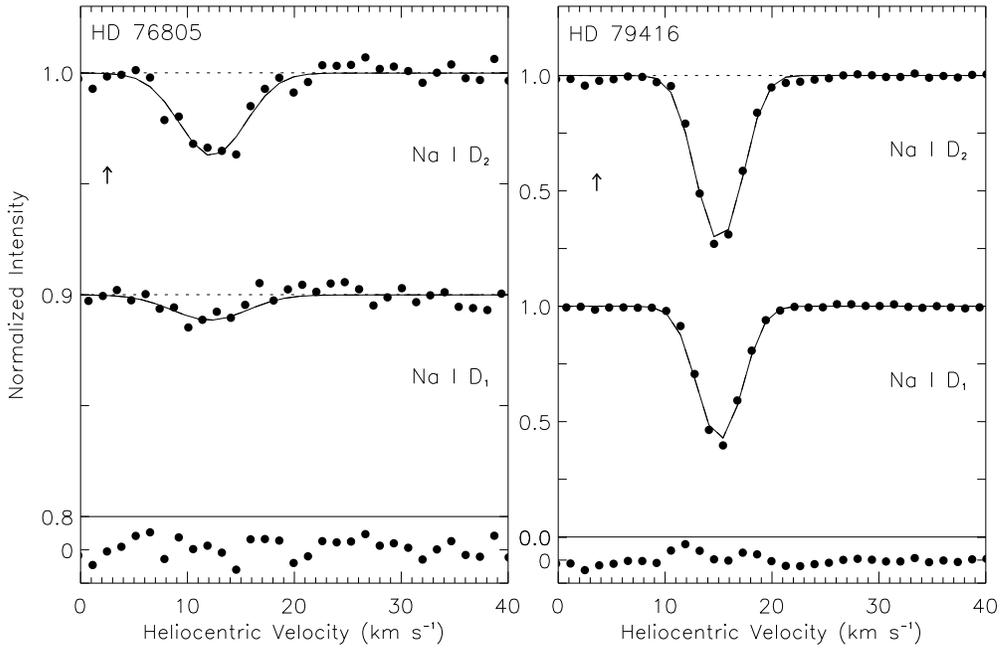
Figure~2.  \ion{Na}{1} D$_2$ and D$_1$ spectra of 11 sight lines towards 
Puppis-Vela with d $<$ 200 pc.  Normalized intensity is plotted versus 
heliocentric velocity (\kms) for each star. The best-fit models to the
absorption line profiles are shown as solid lines and the individual
component fits are indicated by dashed lines.  The bottom panel for
each line of sight shows the residuals to the fits with the y-axis 
ranging from $-$0.1 to +0.1 for HD\,72232 and HD\,79416, and the
y-axis ranging from $-$0.015 to +0.015 for all other lines of sight.
The arrow on each plot
indicates the projected LIC velocity for the particular sight line.
$\dagger$The complete \ion{Na}{1} spectra toward HD\,62226 and HD\,74146
are shown in Figure~6.

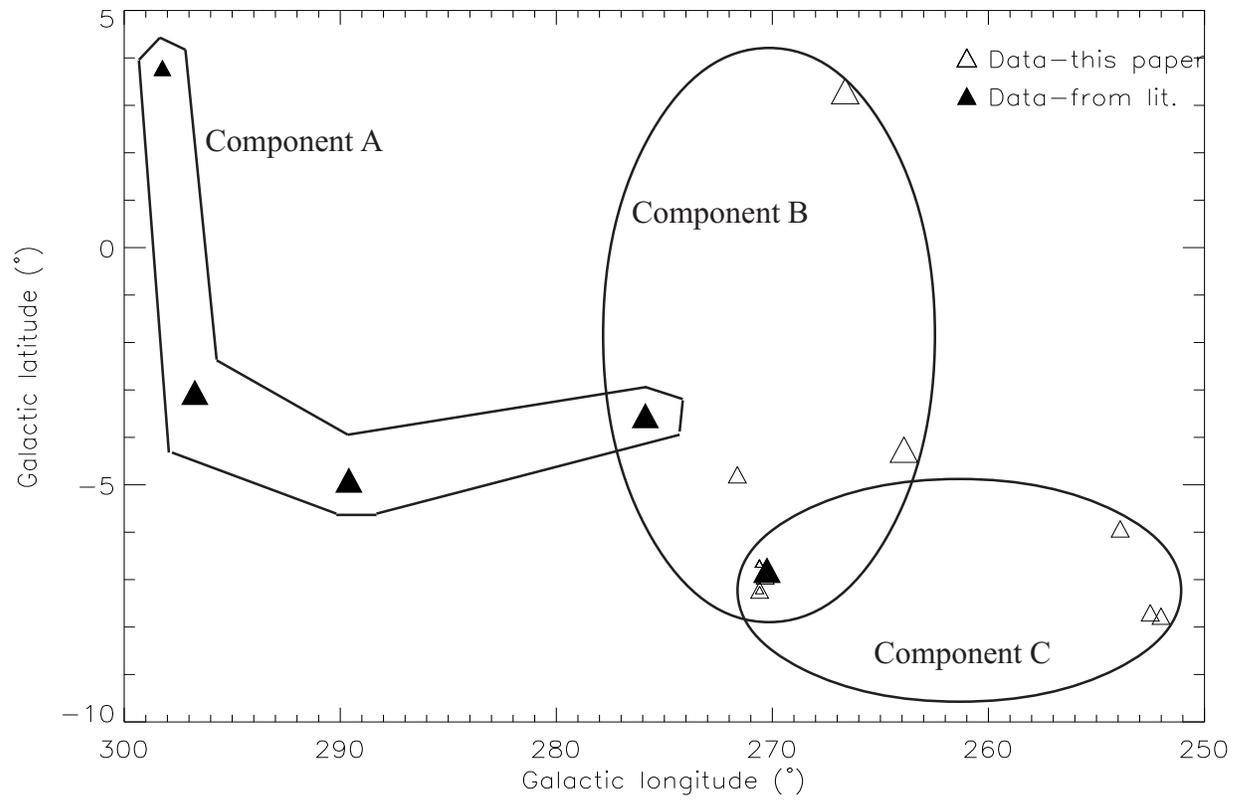
Figure~3.  Panel (a) shows the Galactic positions of the LISM stars towards
the Puppis-Vela for which \ion{Na}{1} spectra were obtained (filled
symbols) and those for which \ion{Na}{1} kinematic data was found in
the literature (open symbols).  Only stars at Galactic latitudes
of $-$10\degr\ $<$ $b$ $<$ $+$5\degr\ have been included.  
Panel (b) shows where \ion{Na}{1} was detected
with velocities of +6 to +9 \kms, (c) shows where velocities of +12 to 
+15 \kms\
were seen, and (d) gives the locations of +21 to +23 \kms\ \ion{Na}{1} gas.  
In panels (b)-(d), the relative symbol size indicates the \ion{Na}{1} column 
density.  Small symbol: 10.00 $\le$ log N(\ion{Na}{1}) $<$ 10.50. 
Medium symbol:
10.50 $\le$ log N(\ion{Na}{1}) $\le$ 11.10.  
Large symbol: 11.70 $\le$ log N(\ion{Na}{1}) $\le$ 12.60.

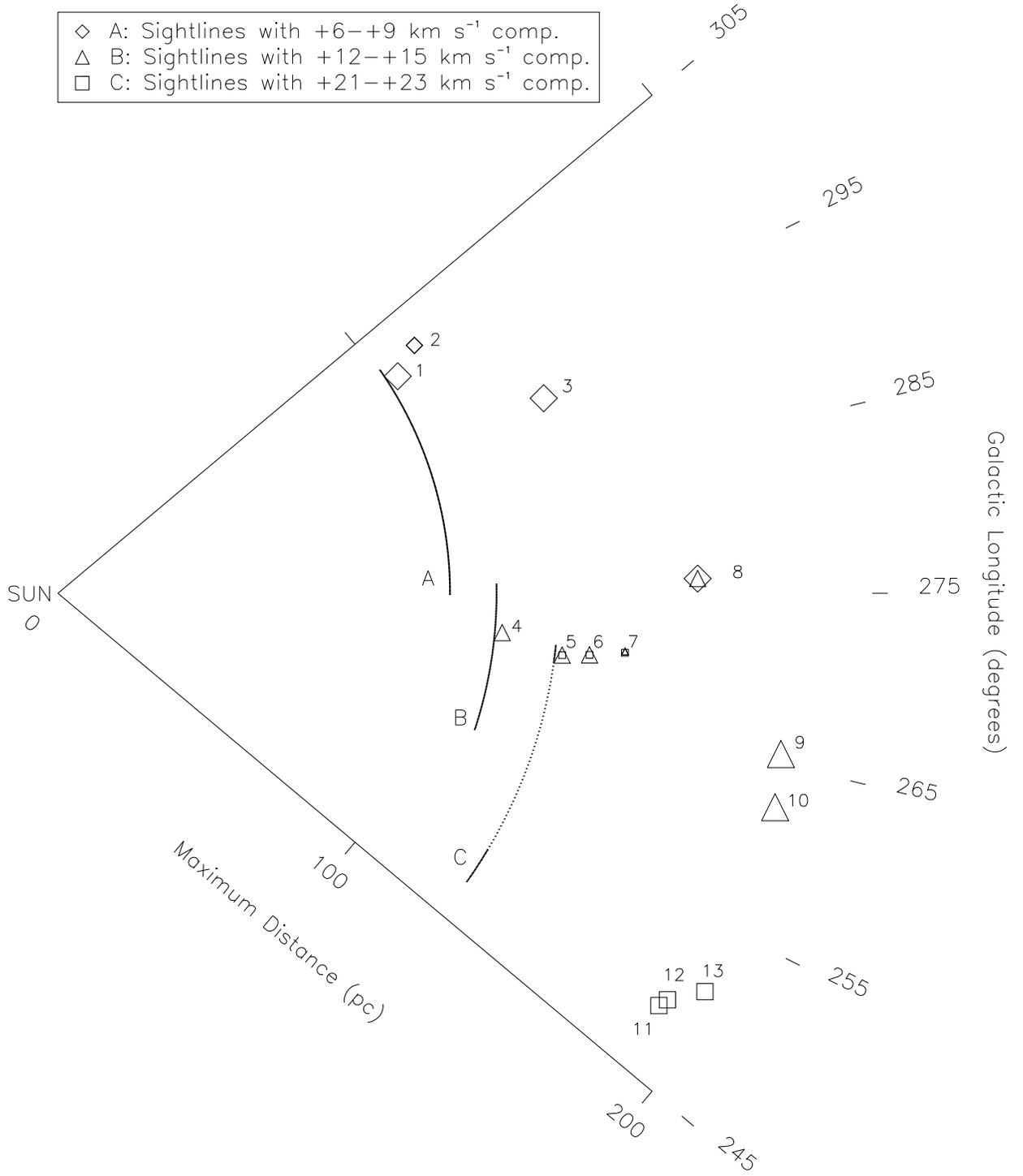
Figure~4.  The LISM stars in Puppis-Vela for which
there exists \ion{Na}{1} kinematic data, projected onto the Galactic plane.
The distance to the stars increases radially, and the Galactic longitude 
increases along the arc.  All stars
have Galactic latitudes of -10\degr$<=$ $b$ $<=$+5\degr.  The diamonds
denote the sight lines in which (A) +6 $<=$ V$_{helio}$ $<=$ +9 \kms\ 
\ion{Na}{1}
gas was detected, the triangles and squares mark where velocities of (B)
+12 $<=$ V$_{helio}$ $<=$ +15 \kms\ and (C) +21 $<=$ V$_{helio}$ $<=$ 23 \kms\
\ion{Na}{1} gas, respectively, were seen.  
The symbol sizes depict N(\ion{Na}{1}) as defined
for Figure~3.  The arcs are drawn to indicate the maximum distance,
and minimum extent in Galactic longitude of gas pockets with velocities
A, B, and C.  The stars shown are: 1) HD\,103079, 2) HD\,106490, 3) HD\,93030,
4) HD\,76805, 5) HD\,74146, 6) HD\,74071, 7) HD\,74560,
8) HD\,81188, 9) HD\,79416, 10) HD\,72232, 11) HD\,61878, 12) HD\,62226,
13) HD\,64503.

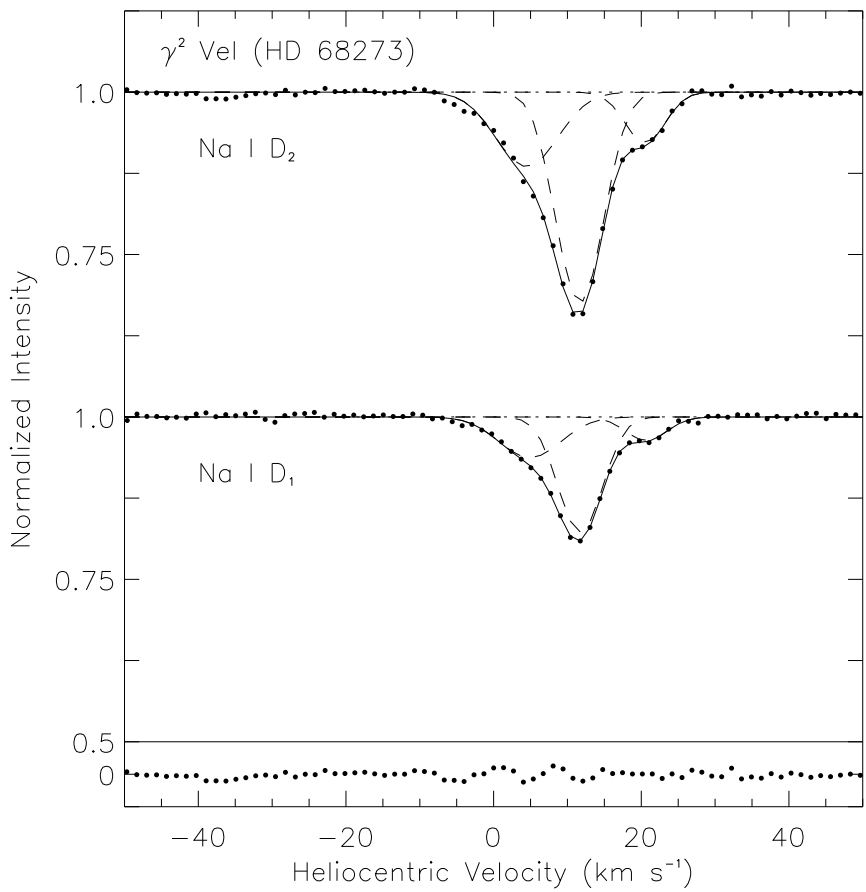
Figure~5.  \ion{Na}{1} D$_2$ (top) and D$_1$ (middle) spectra along the line 
of sight to $\gamma$$^2$ Vel (HD\,68273).  The data points from each spectrum
are represented as dots, a solid line indicates the best-fit model, and the
dashed lines show the three components that compose the fit.  The residual
to the D$_2$ line fit is plotted at the bottom of the panel where the y-axis
ranges between $\pm$0.05. 

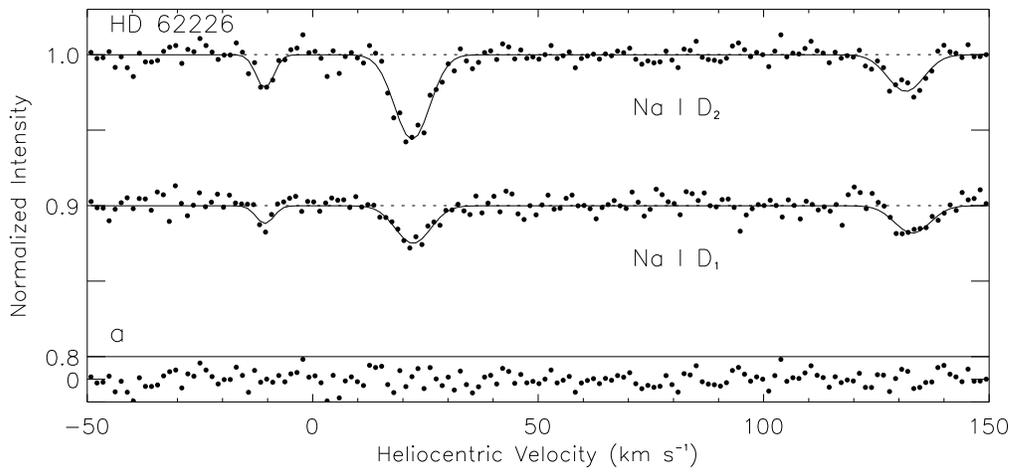
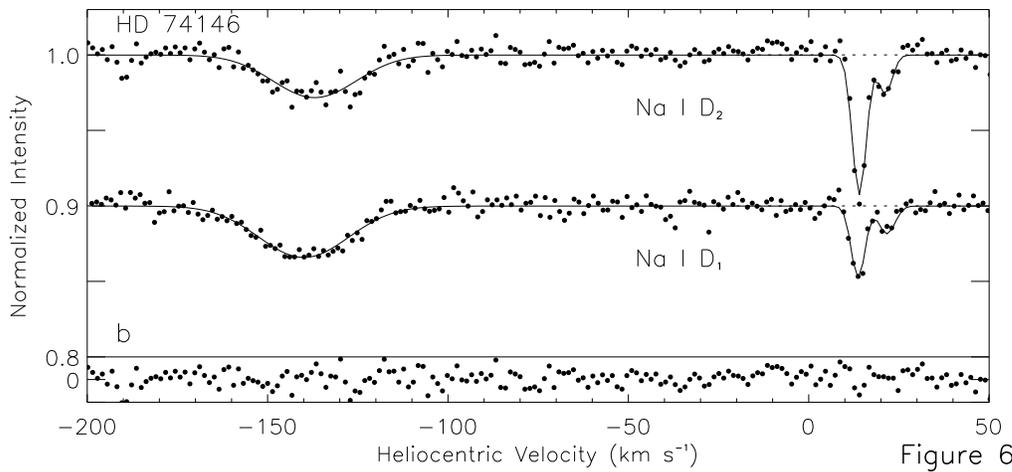
Figure~6.  Complete spectra for the sight lines to a) HD\,62226 and 
b) HD\,74146.  In each panel, normalized intensity of the \ion{Na}{1}
D$_2$ lines are on top and D$_1$ lines are in the middle.  At the bottom
of each panel the residual to the D$_2$ model fits are shown
with the y-axes ranging from $-$0.015 to +0.015.  An IUE spectrum of HD\,62226
was analyzed to help determine the origin of the high velocity components,
but was inconclusive due to the low signal-to-noise ratio.

Figure~7.  \ion{Na}{1} absorption line
data in the direction of the $\beta$ CMa tunnel.  
The solid lines confine the first 200 pc of the
$\beta$ CMa tunnel as proposed by Welsh (1991) and the dotted lines outline
the first 200 pc of the tunnel described in Welsh {\it et al.} (1998).
The thick arc highlights the longitudinal extent of the $\beta$ CMa 
tunnel according to the compilation of the data and pertains to 
the region below the Galactic plane where $-$30\degr $<$ $b$ $<$ $-$10\degr.
The arrows indicate that it is a lower limit on the distance to the end of the
tunnel.
Symbol sizes indicate the column density of \ion{Na}{1}
observed for given lines of sight as defined for Figure~3.  If only
an upper limit on the column density for a sight line is known, a
downward triangle is used for the data point rather than a diamond.
Filled symbols correspond to stars with $-$10\degr\ $<$ $b$ $<$ 0\degr, and
open symbols are used for stars with $-$30\degr\ $<$ $b$ $<$ $-$10\degr.

\end{document}